**Experimental assessment of the speed of light perturbation in free-fall absolute gravimeters**

H. Baumann°, F. Pythoud°, D. Blas*, S. Sibiryakov*$^{\Sigma\Omega}$, A. Eichenberger°, E. E. Klingelé $^\Delta$.

| | |
|---|---|
| ° | Federal Office of Metrology METAS, CH-3003 Bern-Wabern, Switzerland. E-mail: henri.baumann@metas.ch |
| * | Theory Division, CERN, CH-1211 Geneva 23, Switzerland |
| Ω | FSB/ITP/LPPC, Ecole Polytechnique Fédérale de Lausanne, CH-1015 Lausanne, Switzerland |
| Σ | Institute for Nuclear Research of the Russian Academy of Science. 60[th] October Anniversary Prospect, 7a, 117312 Moscow, Russia |
| Δ | Gravity Consulting, Lerchenberg 4, CH 8046 Zurich, Switzerland |

**Abstract**

Precision absolute gravity measurements are growing in importance, especially in the context of the new definition of the kilogram. For the case of free-fall absolute gravimeters with a Michelson-type interferometer tracking the position of a free falling body, one of the effects that needs to be taken into account is the 'speed of light perturbation' due to the finite speed of propagation of light. This effect has been extensively discussed in the past, and there is at present a disagreement between different studies. In this work, we present the analysis of new data and confirm the result expected from the theoretical analysis applied nowadays in free-fall gravimeters. We also review the standard derivations of this effect (by using phase shift or Doppler effect arguments) and show their equivalence.

# 1 INTRODUCTION

The accurate determination of the local acceleration due to gravity, *g*, is important in many different scientific areas like geodesy or geophysics. In metrology *g* plays a crucial role in the so-called Watt balance experiment, aiming at a new definition of the kilogram [1].

The most widely used technique for precision measurement of the gravitational acceleration is to track the position of a free-falling body by means of a Michelson interferometer. The relative uncertainty of this kind of instruments is nowadays a few parts in $10^9$ [2]. This is smaller than the perturbation coming from the finite speed of propagation of light in the set-up[1]. Thus, the latter (known as 'speed of light correction') should be taken into account to properly identify the local gravitational acceleration with current data. The 'speed of light correction' was computed in several works in the past [3-11]. A systematic derivation and comparison of the different previous results (with their limitations) were presented recently [12, 13].

These results were questioned by the authors of [15], who proposed a reconsideration of the theoretical derivation that deviates from [12] (and previous works [3-11]). The controversy was deepened after the claim in [16] of the first measurement of the 'speed of light perturbation' that apparently confirmed the calculations in [15]. The results and methods of [15,16] were questioned in a series of works [17,18] (see also [19]) as well as more recently in [20].

The above controversy, which also has a direct impact on the watt balance experiments [21,22], calls for an independent theoretical analysis and experimental study of the 'speed of light perturbation'. This is the idea behind the present work. We present here the analysis of new sets of data from different gravimeters and find experimental agreement with the classical result given in [12,3-11]. We also review the derivation of the 'speed of light perturbation' by two different methods and show their mutual agreement and equivalence to the results in [12,3-11].

The paper is organized as follows: in section 2 we review the derivation of the 'speed of light perturbation' in Michelson interferometers with two different approaches and show that both calculations coincide. The results of section 2 are applied for the two different cases in section 3: the constant velocity and free falling moving mirror. The latter is the one relevant for gravimetry. Finally, in section 4, we perform the data analysis with an extensive uncertainty evaluation and present a measurement of the effect of the 'speed of light perturbation'. We summarize our results in conclusions. Further descriptions of the data can be found in the annex.

# 2 Michelson interferometer with a moving and a reference mirror

As mentioned in the Introduction, it seems worth to carefully review the derivation of the 'speed of light perturbation' in free-fall gravimeters with a Michelson interferometer. We will present two different derivations (by using the phase shift analysis or by double Doppler effect) and then prove their equivalence. A large part of our analysis can be found in earlier references, e.g. [12, 3-11].

## 2.1 Displacement measurements with Michelson interferometers

The working principle of displacement measuring systems based on interferometry is schematically represented in Figure 1. It is described using a two dimensional reference system with coordinates y and z.

---

[1] As described in [14], these perturbations do not test the foundations of special relativity, and are just related to the finite speed of propagation of light.

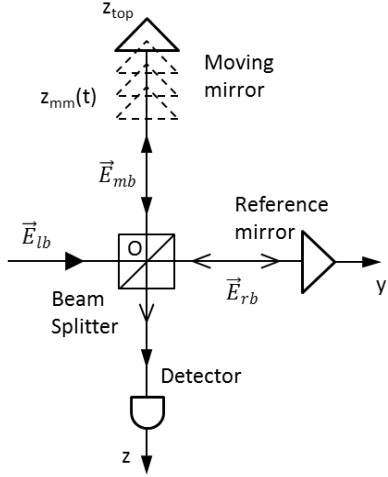

Figure 1: The laser beam, $\vec{E}_{lb}$, is split into the *measuring* beam, $\vec{E}_{mb}$, reflected by the moving mirror, and the *reference* beam, $\vec{E}_{rb}$, reflected by the reference mirror. The two reflected beams are then recombined and sent to the detector.

The light produced by a laser beam source, that we describe here by an electric field $\vec{E}_{lb}(t, y, z)$, is sent into a beam splitter. The incoming beam at the splitting point *0*, is characterized by the electric field,

$$\vec{E}_{lb}(t, 0) = \vec{E}_{lb,0} e^{-i\omega t}, \quad (1)$$

where $\omega$ corresponds to the frequency of the laser beam and where we assumed that the polarization of the beam $\vec{E}_{lb,0}$ is constant in time. We have identified the origin of coordinates with the splitting point *0*. We define the *y*-axis to coincide with the direction of propagation of the incoming beam. After splitting, the *measuring* beam, $\vec{E}_{mb}$, propagates in a direction perpendicular to *y* and that we identify with the *z* direction. This beam is reflected by a moving mirror with trajectory $z_{mm}(t)$ (the *z* coordinate grows in the direction of propagation of the reflected beam, downwards in Figure 1) and generates an electric field at the origin, $\vec{E}_{mb}(t, 0)$. The rest of the original laser beam (the *reference* beam, $\vec{E}_{rb}$) propagates in the *y* direction, is reflected from a fixed reference mirror placed at distance *L* and produces the field $\vec{E}_{rb}(t, 0)$ at the origin. The reference and the measuring beams are recombined at the origin and directed to a photodiode to be converted in an electrical signal. The relevant observable is the intensity *I* resulting after averaging the total field at the origin over times $T_I$, longer than the inverse frequencies of the beams[2], $T_I \gg \frac{1}{\omega}$, but short enough that the phase differences of the two beams are constant (this corresponds to $T_I \ll c/(\omega \frac{dz_{mm}(t)}{dt})$)

$$I(t) = \langle (\vec{E}_{mb}(t,0) + \vec{E}_{rb}(t,0)) \cdot (\vec{E}^*_{mb}(t,0) + \vec{E}^*_{rb}(t,0)) \rangle_{T_I}. \quad (2)$$

---

[2] This is also valid for fields with a time-varying frequency $\omega$ such that $\frac{d\omega}{dt} \ll \omega/T_I$. This condition will hold in our experimental setup.

This quantity will have an oscillatory behavior coming from the interference of the two waves (see below). Timing and counting the occurrence of the interference fringes underlies one of the methods to reconstruct the trajectory of the moving mirror.

We will now proceed to the computation of the quantities appearing in (2) by two standard methods and show their equivalence.

### 2.2 Method using phase analysis and time delay

This analysis is based on the wave nature of the signals, with the propagation speed *c*. To determine $\vec{E}_{mb}(t, 0)$, let us trace the measuring beam backwards in time. The signal at the interference point propagated after reflection from the falling mirror at position

$$z_{mm}(t_r) = -c\, \Delta t(t), \quad (3)$$

where the previous expression defines $\Delta t(t)$ after introducing the reflection time

$$t_r \equiv t - \Delta t(t) . \quad (4)$$

We have made explicit that the variable $\Delta t$ depends on time. Assuming that the reflection is instantaneous, the relation between the reflected waveform and the incident one is given by a phase shift by a constant $\varphi_r$, close to $\pi$ for an ideal mirror. Finally, the incident wave was originated after splitting at time: $t - 2\Delta t$. Following the previous logic, one writes

$$\vec{E}_{mb}(t,0) = \vec{E}_{mb}(t - \Delta t, -c\Delta t) = e^{i\varphi_r}\vec{E}_{ib}(t - \Delta t, -c\Delta t) = e^{i\varphi_r}\vec{E}_{ib}(t - 2\Delta t, 0) \; , \quad (5)$$

where $\vec{E}_{ib}$ is the electric field of the incident beam (the beam after the original laser beam splits).
A similar expression holds for the reference beam, but this time with a fixed distance *L* and with a different incident beam[3],

$$\vec{E}_{rb}(t,0) = e^{i\varphi_r}\vec{E}_{ib}(t - 2L/c, 0). \quad (6)$$

Since both incident beams come from the splitting of the original signal, they share the same phase with the incoming beam $\vec{E}_{lb}$ in (1) except for an irrelevant constant.
The previous formulae (5) and (6) are very generic. In the case of an optical interferometer, one can recall from (1) that we are dealing with monochromatic electromagnetic waves. The two beams produced after splitting will be waves with the same frequency. Applying the previous formulae we then find

$$\vec{E}_{mb}(t,0) = e^{i\varphi_r}\vec{E}_{ib,0} e^{-i\omega(t-2\Delta t)} \,, \qquad \vec{E}_{rb}(t,0) = e^{i\varphi_r}\vec{E}_{ib,0} e^{-i\omega(t-2L/c)}. \quad (7)$$

The intensity is given by

---

[3] For simplicity we assume that this phase is the same as in the measuring beam. Any other constant shift would not change our result.

$$I(t) = I_0 \cdot cos^2[\omega \cdot (\Delta t(t) - L/c)] = I_0 \cdot cos^2[(z_{mm}(t_r) + L)\omega/c], \quad (8)$$

where $I_0$ is a constant irrelevant for the analysis of the trajectory. From equation (8) it can be deduced that the interference fringes occur every time the function $z_{mm}(t_r)$ changes by $\lambda/2$, where $\lambda = \frac{2\pi c}{\omega}$. As mentioned above, one also realizes that by timing and counting the occurrence of the interference fringes, the position of the moving prism at the time when the reflection happens, as seen by the detector, can be determined as a function of time. We also want to remind that the previous formula is just based on the constant speed of light $c$ and is not testing any other aspect of special relativity [14].

### 2.3 Method using Doppler shift and equivalence of the approaches

The previous derivation can be found in many classical textbooks, e.g. [23]. The relevant point is that the intensity function depends only on the variable $\Delta t$ that corresponds to the time needed by the light to travel from the moving mirror to the origin, where the interference occurs. The intensity function is independent of any Doppler shifted frequency. Still, another method to compute the same observable is based on considerations involving the Doppler shift. Since part of the discrepancy in the works [12] and [15] came from using this second method, it seems relevant to derive again the previous result and show that both methods agree. For this, let us consider the electric field of the beam reflected from the mirror as a function of time. The wave has a time varying frequency $\widetilde{\omega}_{mb}$ such that the electric field reads[4]

$$\vec{E}_{mb}(t, 0) = e^{i\varphi_r} \vec{E}_{ib,0} e^{-i \int_{t_0}^{t} d\tau\, \widetilde{\omega}_{mb}(\tau)} \quad . \quad (9)$$

To compute the time dependent frequency, one recalls that this is generated by the absorption and emission of the wave by the moving mirror, which happens at the reflection time $t_r$ defined in (2). This yields the double Doppler shift of the frequency

$$\widetilde{\omega}_{mb} = \left(\frac{1+v(t_r)/c}{1-v(t_r)/c}\right)\omega \,, \quad (10)$$

where $v(t_r)$ is the velocity of the moving mirror at the time of reflection. To show the equivalence between the equations (9) and (7), it is important to realize that the velocity in the previous formula corresponds to the variation of the position with respect to the reflection time

$$v(t_r) = \frac{dz_{mm}}{dt_r} = -c\left(\frac{dt}{dt_r} - 1\right) \,,$$

where in the second equality we used (4). This gives

$$\int_{t_0}^{t} d\tau\, \widetilde{\omega}_{mb} = \omega(2t_r(t) - t) = \omega(t + 2z_{mm}(t_r)/c) \,,$$

and expressions (9) and (7) coincide for any trajectory $z_{mm}(t)$.

It seems interesting to understand why the double Doppler shift did not appear in the derivation of the previous section.

---
[4] In this formalism the constant phase shifts can also be absorbed in the definition of $t_0$.

For this, let us reconsider the intermediate steps in equation (5). Indeed, after reflection by the moving mirror, the frequency of the measuring beam that we observe at time $t$ changed to the value $\widetilde{\omega}_{mb}$. Then equation (5) implies

$$\vec{E}_{mb,0} e^{-i\widetilde{\omega}_{mb}(t_r - z_{mm}(t_r)/c)} = e^{i\varphi_r} \vec{E}_{ib,0} e^{-i\omega(t_r + z_{mm}(t_r)/c)} \,.$$

From the previous expression we can find the value of $\vec{E}_{mb,0}$. Then we find for the field at time $t$ and at the origin,

$$\vec{E}_{mb}(t, 0) = \vec{E}_{mb,0} e^{-i\widetilde{\omega}_{mb} t} =$$
$$e^{i\varphi_r} \vec{E}_{ib,0} e^{-i\omega(t_r + z_{mm}(t_r)/c) - i\widetilde{\omega}_{mb} t + i\widetilde{\omega}_{mb}(t_r - z_{mm}(t_r)/c)} =$$
$$e^{i\varphi_r} \vec{E}_{ib,0} e^{-i\omega(t + 2z_{mm}(t_r)/c)},$$

where in the last equality we used $z_{mm}(t_r)/c = t_r - t$. Thus, the final expression simply depends on the phase change along the light ray (as expected) and is strictly identical to equation (7).

## 3 Two cases: constant velocity and free-falling mirror

In this section, we will apply the general formula (8) to two special cases of motion of the moving mirror: constant velocity and free-fall in the gravity field of the Earth. The latter is of immediate relevance for the experimental analysis in the subsequent section.

### 3.1 Constant velocity

Let us first consider the case of a mirror moving along the $z$ axes from an initial position $z_0$ at a constant velocity $v_0$,

$$z_{mm}(t) = z_0 + v_0 \cdot t.$$

From the definition of $\Delta t$, equation (3), we find,

$$\Delta t = \frac{z_0 + v_0 \cdot t}{v_0 - c}.$$

This eventually translates into the intensity of the form (cf. (8))

$$I = I_0 \cdot cos^2\left[\omega \cdot \left(\frac{1}{c - v_0}(z_0 + v_0 \cdot t) + L/c\right)\right].$$

As we know from the general treatment, the trajectory dependent term corresponds to the time of reflection of the wave

$$z_{mm}(t - \Delta t) = \frac{1}{1 - v_0/c}(z_0 + v_0 \cdot t).$$

In this last equation we see the correction factor $1/(1 - v_0/c)$, that corresponds to the double Doppler shift also discussed in [15].

### 3.2 Constant acceleration

Let us now turn to a freely falling mirror in the gravitational field of the Earth. For our analysis we will consider the approximation, in which the latter produces an acceleration characterized by a constant term and a gradient term,

$$\frac{d^2 z_{mm}(t)}{dt^2} = g + \Gamma\, z_{mm}(t) \quad . \quad (11)$$

Inclusion of the vertical gravity gradient is necessary to achieve the relative precision of a few parts in $10^9$ in the measurements of $g$. Indeed, the measured values of $\Gamma$ are in the range of 3 $\mu$Gal/cm (1 $\mu$Gal = $10^{-8}$ m/s$^2$) with an associated relative uncertainty of 3% (k=2)[5]. In the previous formula $g$ is the acceleration at the origin of coordinates. We omit other sources of disturbances, such as self-attraction, in this theoretical treatment, as they essentially depend on the properties of the experimental apparatus. Their discussion is postponed to the next section.

After integration equation (11) and expanding in $\Gamma t^2$ one finds

$z_{mm}(t) = z_0(1 + \Gamma t^2/2) + v_0 t(1 + \Gamma t^2/6) + g t^2/2 (1 + \Gamma t^2/12)$   . (12)

Even if the previous equation yields a quartic equation for $\Delta t(t)$, by noticing that $\Delta t(t)$ is of order $z_{mm}(t)/c$ one can reduce the calculation of the first 'speed of light' perturbation to the solution of a quadratic equation. Neglecting the contributions of $O(c^{-3})$ and $O(\Gamma c^{-2})$, one finds

$\Delta t = -[z_0 + v_0 t + (g + \Gamma z_0)t^2/2 + 1/6\, v_0 \Gamma t^3 + 1/24\, g\, \Gamma\, t^4]/c - [v_0 z_0 + (v_0^2 + g z_0)t + 3/2\, gv_0 t^2 + g^2 t^3/2]/c^2$ .

Finally, this generates a time dependent intensity (8) characterized by the function

$z_{mm}(t_r) = z_{mm}(t - \Delta t) = (z_0 + v_0 t + gt^2/2) + (v_0 z_0 + (v_0^2 + g z_0)t + 3/2\, gv_0 t^2 + g^2 t^3/2)/c + \Delta z^\Gamma(t)$,   (13)

where the piece depending on the gravitational gradient reads

$\Delta z^\Gamma(t) \equiv \Gamma t^2/2\, (z_0 + v_0 t/3 + 1/12\, g\, t^2)$ .

In equation (13) we see how the perturbation due to the finite value of *c* enters into the final formula. Notice that those do not affect the contribution from the gradient at the order of interest.

For gravimetry, one needs the connection between the time variations of the intensity and the value of the local gravitational acceleration *g*. This follows simply from taking the second derivative of the variable inside the cosine in (8). At the desired accuracy one gets

$\frac{d^2}{dt^2} z_{mm}(t - \Delta t) = g + 3 \cdot \left(\frac{g \cdot v_0 + g^2 \cdot t}{c}\right) + \Delta g^\Gamma(t) = g + g_{bias}(t) + \Delta g^\Gamma(t)$ .     (14)

In the previous expression we have denoted by $\Delta g^\Gamma(t)$ the contribution coming from the gradient and introduced the quantity related to the speed of light by

$g_{bias}(t) \equiv 3 \cdot \left(\frac{g \cdot v_0 + g^2 \cdot t}{c}\right)$ .    (15)

This coincides with the standard result revised in [12].

---

[5] The standard uncertainty, or uncertainty *u*, of a result of measurement reflects the lack of exact knowledge of the value of the measurand. The expanded uncertainty, denoted by *U*, is obtained by multiplying the uncertainty by a *coverage factor k* [29].

Our aim is to verify the formula (15) for the speed of light perturbation experimentally. To be open to possible experimental surprises, we will treat the multiplication factor on the right side of (15) as a free parameter in the analysis of the data and replace (15) by the function

$g_{bias}(a_c, t) \equiv a_c \cdot \left(\frac{g \cdot v_0 + g^2 \cdot t}{c}\right)$.  (16)

The result derived in [15] corresponds to $a_c = 2$. We believe that this is due to an error in the analysis of [15] where it was not taken into account that the Doppler change in frequency (which occurs at reflection) and the interference at the beam splitter happen at different times.

## 4    Experimental study

The purpose of our study is to determine experimentally the proportionality factor $a_c$ defined in (16). We used the data acquired by different free fall absolute gravimeters at different sites. The working principle of this kind of devices has been widely described elsewhere [2]. In these instruments, the trajectory of a corner cube, free falling in the Earth gravity field, is measured with a Michelson type interferometer.

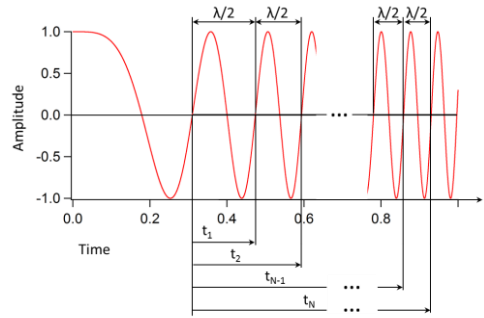

Figure 2: Illustration of the interference pattern measured by the photodiode. By timing and counting the occurrence of the fringes (zero crossings after mean-value subtraction in this example), position-time pairs $(z_i; t_i)$ are determined, and the position of the free falling body as a function of time can be estimated. The signal is normalized to lie in the interval$(-1,1)$.

To illustrate the measuring principle of a free-fall absolute gravimeter through the time dependence of the intensity, equation (8), we show an expected interference pattern in Figure 2. From (8) and (13) one deduces that if there were no speed-of-light perturbation, the interference fringes would occur each time the mirror moves by a distance $\lambda/2$. Hence the shrinking of the spacing between the fringes for accelerated trajectories. Note, however, that the 'speed of light perturbation' in (13) modifies this relation between the position of the fringes and the trajectory of the mirror.

By measuring the times at which the intensity (8) goes through the mean value, $t_i$ (in the case illustrated by Figure 2 the mean value is shifted to zero), we can associate this to a zero of $I(t)$ and determine the set of points $z_i$ (up to an irrelevant constant). These position-time pairs $(z_i; t_i)$ allow reconstructing the function $z_{mm}(t_r(t))$ appearing in (8)
.

## 4.1 Experimental description

To determine the proportionality factor $a_c$ in (16), we use the same method as Rothleitner in [16]. This method is based on ignoring in (13) the contribution coming from the speed of light perturbation and fitting the data by the formula obtained in the limit $c \to \infty$. This leads to the following model function

$$z_{mm}(t_r) = z_{mm}(t - \Delta t) = (z_0 + v_0 t + g t^2/2) + \Delta z^\Gamma(t), \quad (17)$$

to which the data of each drop were fitted using the least squares method.

Since the gravity gradient is known from other experiment (cf. below) the fit produces experimental values for $z_0$, $v_0$ and $g$. The value of $g$ measured this way, that we call $g_{meas}$, will produce a value varying as a function of the total drop time $T$ and the initial velocity $v_0$. The study of this variation allows one to determine $a_c$.

The applied procedure is illustrated by Figure 3. In our experiment, typically a few 1000 position-time pairs $(z_i; t_i)$ were acquired for a single drop. By removing the fringes at the end of the drop (*Last fringes removal*) the obtained values of $g_{meas}(T, v_0)$ will all have the same initial velocity $v_0$ but their total drop time $T$, will be different. By removing fringes at the beginning of the drop (*First fringes removal*) the obtained values $g_{meas}(T, v_0)$ will depend both on the total drop time $T$ and the initial velocity $v_0$.

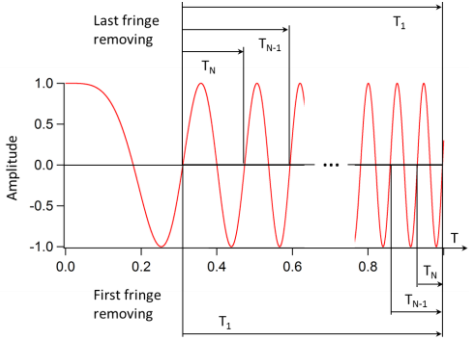

Figure 3: Schematic representation of the modulation of the total drop time $T_N$ considered for the determination of $g_{meas}(T_N, v_{0N})$. By removing the fringes at the end of the drop (*Last fringe removal*), the obtained values for *g* corresponds to the same initial velocity $v_0$. They will only differ because of different drop times $T_N$. On the other hand, when fringes from the beginning of the drop are removed (*First fringe removal*) the obtained values for *g* depend on the different drop times $T_N$ and the different initial velocities $v_{0N}$.

The number of drops processed per set (a set is given by a number of drops acquired at a given date on a given station) is varying between 2000 and 5000 with a typical drop interval of 10 seconds. To single out the speed of light perturbation, and exclude any influence from time dependent gravity variations, all experimental values $g_{meas}(T, v_0)$ were corrected for all classical geophysical perturbations, as well as for the self-attraction [24] and transferred to the same height. We emphasize that the perturbation due to the finite speed of light was not included into the fitted model (17). To focus on the speed of light perturbation in the measured value of the gravitational acceleration, we subtracted from $g_{meas}(T, v_0)$ the minimum value of the set to obtain $\Delta g_{meas}(T, v_0)$. These values were then least square fitted to

$$\Delta g_{meas}(T, v_0) = a_c \cdot \Delta g_{theo}(g, T, v_0), \quad (18)$$

with

$\Delta g_{theo}(g, T, v_0) = \left(g \cdot v_0 + 0.5 \cdot g^2 \cdot T \cdot (1 + \eta_2/5)\right)/c$, and where $\eta_2 \equiv 5 \cdot \lambda_b \cdot \left(\lambda_b^2 - 3\right) / \left(7 \cdot \left(3 \cdot \lambda_b^2 - 5\right)\right)$, with $\lambda_b \equiv 1/(1 + 2 \cdot v_0/(g \cdot T))$ are introduced to take in consideration that the data acquired by the gravimeter are equally spaced in distance [11].

The impact of the speed of light perturbation on the estimation of *g* has been summarized in [11,15]. Considering a total drop time $T = 0.2\ s$ and an initial velocity $v_0 = 0.2\ m/s$ the correction due to the finite speed of light can vary from 11 µGal for $a_c = 3$ to 7 µGal for $a_c = 2$.

## 4.2 Experimental results

In this chapter, the results obtained by analyzing data from three different instruments that have been setup on 9 different stations are presented.

Instruments:
- Gravimeter FG5X-209 from the Federal Institute of Metrology METAS.
- Gravimeter FG5X-311 from Micro-g Lacoste.
- Gravimeter FG5X-216 from the University of Luxembourg.

Stations:
- *WANA*: Absolute reference station at METAS.
- *Zimm*: Station of the Swiss geodetic gravity reference network.
- *A1, A3, B3, B5, C3, C4*: Reference stations in the underground laboratory in Walferdange, Luxembourg.
- *Rochfort*: Geodetic reference station Luxembourg.

### 4.2.1 Uncertainty evaluation

The evaluation of the uncertainty associated to $a_c$ is made in two steps. First the uncertainties related to $\Delta g_{theo}$ and $\Delta g_{meas}$ are estimated. Then the uncertainty for $a_c$, obtained by least square on (18), is calculated.

*Uncertainty evaluation for $\Delta g_{theo}$ and $\Delta g_{meas}$*

The uncertainty associated to $\Delta g_{theo}$ is estimated by applying the law of propagation as described in [29]. The results of this evaluation are summarized in Table 1.

|  | $x_i$ / unit | | $\delta(x_i)$ / unit | | $u(x_i)$ / unit | |
|---|---|---|---|---|---|---|
| $v_0$ | 0.20 | m/s | 0.001 | m/s | 0.003 | µGal |
| $g$ | 9.81 | m/s² | 0.001 | m/s² | 0.001 | µGal |
| $T$ | 0.15 | s | 0.001 | s | 0.032 | µGal |
| $u_{\Delta g_{theo}}$ | | | | | *0.032* | *µGal* |

Table 1: Uncertainty budget associated to $\Delta g_{theo}(g, T, v_0)$. '$x_i$ / unit' represents the parameters and its unit, $\delta(x_i)$ the uncertainty associated the to the parameter $x_i$, $u(x_i)$ the contribution of the parameter $x_i$ to the uncertainty $u_{\Delta g_{theo}}$.

The result given in Table 1 shows clearly that even with a very conservative evaluation, the uncertainty associated to $\Delta g_{theo}(g, T, v_0)$ stays very small.

For the evaluation of the uncertainty associated to $\Delta g_{meas}$ it is important to remind that in the present experiment, we are interested in changes of acceleration as a function of the total drop time $T$ and $v_0$ and not in the absolute value. This 'differential' mode eliminates almost all systematic errors. Nevertheless, there are a number of error sources that are dependent of the total drop time $T$ and $v_0$. To estimate their influence we divided them in two groups. The first group contains the sources that we could not include in our model. These are the laser beam diffraction correction, the frequency dependent phase shift in the electronic, the corner cube rotation, the residual air friction and the residual ground vibrations. The contribution of the second group, formed by the self-attraction and the gradient is discussed in a separate paragraph below.

The impact of the first group, without ground vibrations, has been conservatively evaluated to be of order 0.5 $\mu$Gal, in agreement with [16,11,26]. Regarding the vibrations, in the same manner as in [16], we estimated their influence by a spectral analysis of the residual function, where the residual function corresponds to the difference between the model given by equation (17) and the measurements. Figure 4 shows the average amplitude spectrum from the residuals of more than 2500 drops acquired with the instrument 209 on station WANA in the METAS watt balance laboratory.

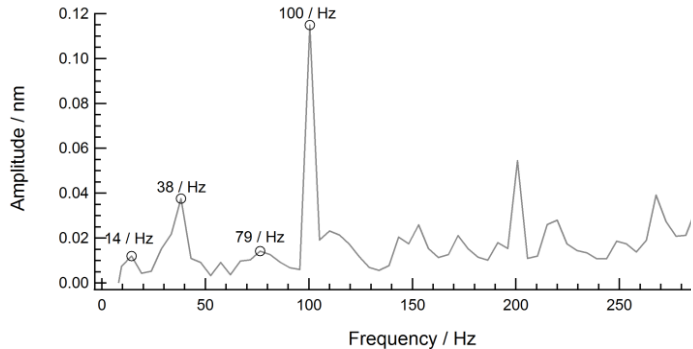

Figure 4: Average amplitude spectrum obtained from more than 2500 drops acquired on station WANA in the watt balance laboratory of METAS with the instrument 209.

The spectrum signature shown in Figure 4, is very similar to those obtained with the other instruments analyzed in the context of the present study, independently of the site where the measurements were made. This indicates that the oscillations are probably generated by the instruments themselves. The error on $g$ induced by instrumental oscillations has been discussed in [27] and more recently in [28]. The oscillation related error is dependent on the amplitude, the frequency, the phase and the total drop time $T$. To evaluate this error, the average amplitude and phase spectrums were estimated for each set. From these spectrums, the error induced by the four main amplitudes in the frequency range between 0 Hz to 150 Hz were estimated by applying the formulas given in [27, 28]. The oscillations dependent error obtained for the different sets of data acquired on station WANA by the instrument 209 are shown in Figure 5. The mean error at the lowest drop time is around 0.3 $\mu$Gal with a standard deviation of 0.5 $\mu$Gal and tends to zero at longer drop times.

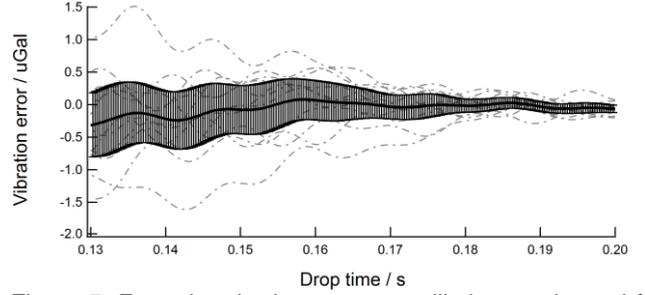

Figure 5: Error due the instrument oscillations estimated for the sets acquired on station WANA by the instrument 209, Each dashed curve corresponds the error estimated for one set. The continuous curve represents the mean value of all individual curves with its associated uncertainty. For a total drop time $T = 0.13$ s the mean error is estimated at around 0.3 $\mu$Gal with a standard deviation of 0.5 $\mu$Gal. The mean error tends to zero with increasing drop time.

Another straight forward and indicative evaluation of the influence of the instrumental oscillations is to convert the amplitude of the main harmonics into acceleration. The estimate obtained by this method is of the order of 0.9 $\mu$Gal. This value is in agreement with the estimate made previously as well as with the estimate given by Rothleitner in [16]. To underpin the different estimations made above, the uncertainty due to the sum of all the sources of the first group, were determined by evaluating $g_{meas}$ as described under 4.1, but including in the model the perturbation due to the finite speed of light (i.e. using the correct formula (13), in that case the theoretical expectation is simply $g$). The standard deviation of $g_{meas}$ was estimated to 1 $\mu$Gal (k=1) which is in agreement with our estimations as well as with those given in [16]. The uncertainty contributions to $\Delta g_{meas}$ are finally summarized in Table 2.

| $x_i$ | $u(x_i)$ / µGal |
|---|---|
| Laser beam diffraction | |
| Electronic phase shift | |
| Corner cube rotation | 0.50 |
| Residual air friction | |
| Standard deviation of the mean | 0.30 |
| Vibrations | 1.00 |
| $u_{\Delta g_{meas}}$ | **1.16** |

Table 2: Uncertainty budget associated to $\Delta g_{meas}$. $x_i$ represents the parameters, $u(x_i)$ the contribution of the parameter $x_i$ to the uncertainty $u_{\Delta g_{meas}}$

*Uncertainty evaluation for $a_c$*

The method used for the estimation of the proportionality factor for one set, is the Levenberg-Marquardt least square orthogonal distance method in which the uncertainties $u_{\Delta g_{meas}}$ and $u_{\Delta g_{theo}}$, associated to $\Delta g_{meas}$ and $\Delta g_{theo}$ respectively, are included in the method. The software IGOR

Pro[6] was used to perform the fit. The uncertainty, $u_{ac\_fit}$, associated to $a_c$ given by the fitting procedure has been estimated to 0.3 (k=1).

As mentioned above, the influence of the self-attraction and the gradient are also dependent of the total drop time. To estimate the contribution of these two factors to $a_c$, a Monte-Carlo simulation was performed. For that, a synthetic data set was generated, taking into account the models of the self-attraction [24] and the gravity gradient [25]. The first and second order parameters of the gravity gradient were varied with a normal distribution and a standard deviation of 5% from the effective value. The self-attraction function was varied around its nominal value with a normal distribution and a standard deviation of 3 %. With 2000 data sets, the uncertainty $u_{ac\_mont}$ in $a_c$, due to gradient and the self-attraction was estimated to be 0.3 (k=1). For this estimation we considered the total drop time, which leads to a conservative evaluation. The uncertainty associated to the proportionality factor $a_c$ estimated for one set is summarized in Table 3.

| $x_i$ | $u(x_i)$ |
|---|---|
| Least square | 0.30 |
| Self-attraction | 0.30 |
| Gradient | |
| $u_{a_c}$ | **0.42** |

Table 3: Uncertainty budget associated to $a_c$. $x_i$ represents the parameters, $u(x_i)$ the contribution of the parameter $x_i$ to the uncertainty $u_{a_c}$

### 4.2.2 Results

*Determination of the proportionality factor $a_c$*

For the experimental determination of the proportionality factor $a_c$ we used the data sets acquired by the three gravimeters, 209, 311 and 316 on the sites WANA, Zimm and Walferdange. As an example, Figure 6 shows $\Delta g_{meas}$ as a function of $\Delta g_{theo}$ for the data sets acquired by the instrument 209 on the station WANA in the watt balance laboratory from METAS.

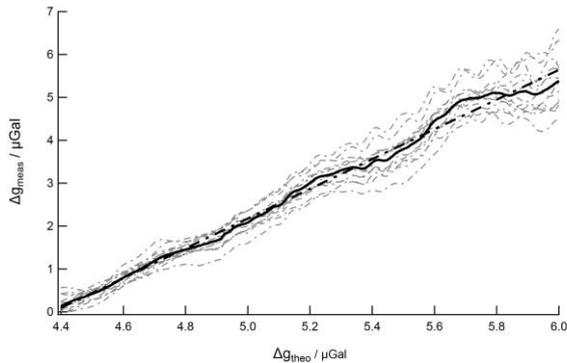

Figure 6: The plot shows $\Delta g_{meas}$ as a function of $\Delta g_{theo}$ for the 13 data sets acquired in the watt balance laboratory on station WANA. The gray dashed curves are representing the

---

[6] IGOR Pro, WaveMetrics Inc., Version 6.21, ODRPACK95.

values obtained for the different sets. The black continuous curve represents the mean of all set curves and the dashed black curve is the linear fit of the mean curve. The slope of that curve corresponds to the mean $a_c$ of all 13 sets. It has been estimated to $3.4 \pm 0.43$ ($k = 1$).

On station WANA, the mean value of the proportionality factors $a_c$ has been estimated to $a_c^{WANA} = 3.4$, with an uncertainty of $0.43$ (k = 1). The contributions to the uncertainty are summarized in Table 4.

| $x_i$ | $u(x_i)$ |
|---|---|
| Least square | 0.30 |
| Self-attraction | 0.30 |
| Gradient | |
| Repeatability, $\sigma_M^{WANA}$ *(standard deviation of the mean of $a_c^{WANA}$)* | 0.08 |
| $u_{a_c}^{WANA}$ | **0.43** |

Table 4: Uncertainty budget associated to $a_c$ estimated on station WANA. $x_i$ represents the parameters, $u(x_i)$ the contribution of the parameter $x_i$ to the uncertainty $u_{a_c}^{WANA}$.

In the same manner as described above, the proportionality factor $a_c^{site}$ has been evaluated for three different instruments at different sites. The results are presented in Figure 7 and summarized in Table 5 (more details of the measurements are given in annex 1).

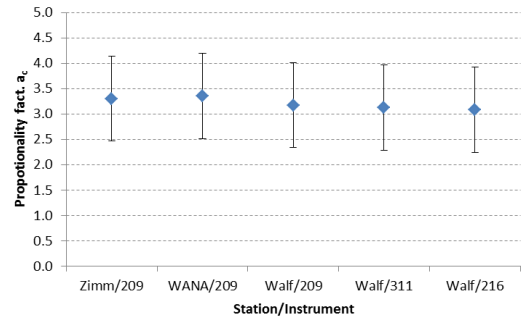

Figure 7: Proportionality factor $a_c^{site}$ for the three instruments, 209, 216 and 311, at sites Zimm, WANA and Walf with their respective associated uncertainty (k=2).

| Station/Instrument | $a_c^{site}$ | $u_{ac}$ |
|---|---|---|
| Zimm/209 | 3.3 | 0.43 |
| WANA/209 | 3.4 | 0.43 |
| Walf/209 | 3.2 | 0.43 |
| Walf/311 | 3.1 | 0.43 |
| Walf/216 | 3.1 | 0.43 |

Table 5: Proportionality factors $a_c^{site}$ and their respective uncertainty estimated for three different instruments at different sites.

Table 5 and Figure 7 shows that the proportionality factors $a_c^{site}$, obtained with the datasets of three different instruments at different sites, are in complete agreement. Based on these five results the mean value of the proportionality factor is estimated to $\bar{a}_c^{site} = 3.2 \pm 0.45$ ($k=1$) where the contributions to the uncertainty are summarized in Table 6.

| $x_i$ | $u(x_i)$ |
|---|---|
| $a_c$ | 0.43 |
| Repeatability $\bar{\sigma}_M^{site}$ (mean of $\sigma_M^{site}$) | 0.10 |
| Reproducibility (standard deviation of the mean of the $a_c^{site}$) | 0.10 |
| $\boldsymbol{u_{\bar{a}_c}^{site}}$ | **0.45** |

Table 6: Uncertainty budget associated to $\bar{a}_c^{site}$ estimated on three different sites. $x_i$ represents the parameters, $u(x_i)$ the contribution of the parameter $x_i$ to the uncertainty $\boldsymbol{u_{\bar{a}_c}^{site}}$.

*Comparison at B3*

The evaluation of the proportionality factor $a_c^{B3}$ presented by Rothleitner [16] has been made with a dataset acquired with the instrument 216 on station B3 in the underground laboratory in Walferdange. In the context of the present study, the dataset used by Rothleitner has been reprocessed with our procedure, together with three other datasets acquired on B3 by the instrument 209 and 216. The obtained results are presented in Figure 8 and summarized in Table 7.

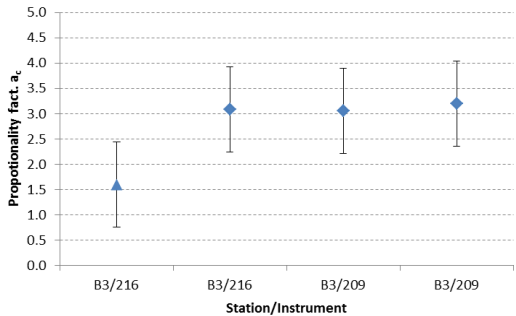

Figure 8: Proportionality factor $a_c^{B3}$ obtained with the data acquired by the instruments 216 and 209 on the station B3 in Walferdange and their respective associated uncertainty (k=2). The point denoted with the symbol ▲ corresponds to the value obtained with the data set acquired on 04.10.12 with the instrument 216. The three point denoted with the symbol ♦ are corresponding to the values obtained on 27.05.14 with 216, and 209 on 06.11.13.

Figure 8 shows that the first estimate of $a_c^{B3}$ by 216 on B3, denoted by the symbol ▲, present a significant offset to the three other estimates. This disagreement is confirmed by the calculation of the compatibility index $E_n$ defined by:

$$E_n = \frac{|x_i - x_{ref}|}{\sqrt{U^2(x_i) + U^2(x_{ref})}},$$

where $x_i$ is the i-th estimation, $x_{ref}$ the estimated reference value and U the respective uncertainties. An $E_n$ factor lager than 1 indicates that the two values are incompatible, as their difference cannot be covered by their uncertainties. It means that either one of the two values is corrupted or the declared uncertainties are too small.

In our case, $x_i$ are the estimated $a_c$ and $x_{ref}$ the weighted mean value of the $a_c$

| Station/Instrument | $a_c^{B3}$ | $u_{a_c}^{B3}$ | $E_n$ |
|---|---|---|---|
| *B3/216 (04.10.12)* | *1.6* | *0.9* | *1.3* |
| B3/216 (27.05.14) | 3.1 | 0.9 | 0.4 |
| B3/209 (06.11.13) | 3.1 | 0.9 | 0.4 |
| B3/209 (06.11.13) | 3.2 | 0.9 | 0.5 |
| Mean_Weighted | 2.8 | 0.4 | |

Table 7: Proportionality factor $a_c^{B3}$ estimated by the instruments 216 and 209 in the underground laboratory in Walferdange on station B3.

The $E_n$ factors from Table 7 indicate that the first estimation of $a_c^{B3}$ (B3/216, 04.10.12), is incompatible with the reference value. After withdrawing the incompatible value we get a weighted mean value $\bar{a}_c^{B3} = 3.1 \pm 0.9$ ($k=2$). This result is in agreement with the result obtained in the previous paragraph.

## 5 Conclusion

In this work, we have established experimentally the value of the proportionality factor used in the speed of light corrections in the position measurements of free-falling mirrors by Michelson interferometers. This is particularly relevant to determine the Earth gravitational acceleration with a relative uncertainty of a few parts in $10^9$. Given the past controversy in the value and origin of this correction we devoted the first two sections to give a thorough review of the effect by two independent methods. Our final theoretical result is that the interference pattern in the intensity of the combined beams measured at the detector of Figure 1 will be characterized by the time dependent pattern (8), where $z_{mm}(t_r)$ is given by (13). In (13) one sees clearly the perturbation related to the finiteness of $c$. This coincides with previous results in the literature, cf. [12] and references therein.

For the experimental confirmation of the $1/c$ perturbation we analyzed 28 datasets, which correspond to more than 50'000 drops, from 3 different instruments on 9 different sites. The results of this analysis agree with the theoretical expectations. They also display a coherent behavior of the instruments, independently of the site or their respective configuration setup. We parameterized the effect (and possible deviations) by the parameter $a_c$ in (16). Our theoretical result is $a_c = 3$ while the experimental analysis yields $\bar{a}_c^{site} = 3.2 \pm 0.9$ ($k=2$) from the combined results of sites Zimm, WANA and Walf, while from the site B3 in Walferdange one gets: $\bar{a}_c^{B3} = 3.1 \pm 0.9$ ($k=2$).

Previous to this work, the perturbation due to the finite speed of light was experimentally assessed by Rothleitner et al. [16]. They obtained a proportionality factor of $a_c = 2$, that coincides with their theoretical expectation derived in [15]. We have reprocessed the data used by Rothleitner in [16] with our own software and have showed that the result obtained with these data is not in agreement with three other results obtained with data acquired on the same site by the same instrument and a second one. Even if we could not identify any clear error in their experimental analysis, we think that our results (with more data and from different stations) hint toward some possible anomaly in their dataset. Concerning the derivation in [15] we think that their analysis does not properly take into account the fact that the interference and reflection times do not coincide.

The precise determination of the Earth gravitational field is a key element of the definition of the kilogram through a Watt-balance experiment [1]. Our results confirm the traditional theoretical treatment of one of the most important corrections to achieve the desired relative uncertainty of a few parts in $10^9$, thus paving the road for the feasibility and reliability of the method.

## 6    Acknowledgments

The authors would like to thanks Dr. Christian Rothleitner, Dr. T. Niebauer and Prof. Olivier Francis for giving us the possibility of reprocessing their dataset and the open discussions. We gratefully thank Vyacheslav Rychkov, Ignatios Antoniadis for their support. Finally, we thank the Micro-g Lacoste team, Dr. D. van Westrum, R. Billson and B. Ellis, for allowing us to include their datasets in our analyses. We also acknowledge the valuable comments and suggestions of the referees of the publication.
This work was partially funded by the European Metrology Research Program (EMRP) participating countries within the European Association of National Metrology Institutes (EURAMET) and the European Union.

**Annex 1: Description of the datasets**

*FG5X 209 on station WANA*

In the context of the Watt-Balance experiment [22], METAS has built a dedicated laboratory in which 5 absolute gravity stations have been defined. From these 5 stations *WANA* is the reference station. At that station, gravitational acceleration is measured approximately once a month since more than 10 years. The standard deviation of the value of $g$ averaged over time at this station is less than 2 $\mu$Gal.

For the present work, 13 datasets, with different number of drops (1500 to 4500), have been processed. The proportionality factors obtained are summarized in Table 8.

| Date | Gravi | Station | $a_c$ |
|---|---|---|---|
| 25.02.2013 | 209 | WANA | 3.1 |
| 27.02.2013 | 209 | WANA | 3.2 |
| 29.04.2013 | 209 | WANA | 3.2 |
| 25.06.2013 | 209 | WANA | 3.3 |
| 02.07.2013 | 209 | WANA | 2.9 |
| 03.07.2013 | 209 | WANA | 3.5 |
| 29.07.2013 | 209 | WANA | 3.4 |
| 04.08.2013 | 209 | WANA | 3.5 |
| 28.08.2013 | 209 | WANA | 3.6 |
| 28.10.2013 | 209 | WANA | 3.1 |
| 04.11.2013 | 209 | WANA | 3.3 |
| 13.11.2013 | 209 | WANA | 3.8 |
| 18.11.2013 | 209 | WANA | 3.8 |
| | | $a_{c\_mean}$ | **3.4** |

Table 8: Proportionality factor evaluated on station WANA with the instrument 209. The value has been estimated at 3.4 with a standard deviation of the mean of 0.08.

*FG5X 209 on station Zimm*

The gravity station *Zimm* is a reference station of the Swiss geodetic gravity reference network. The value of $g$ is measured at that station once a year since about 10 years. The standard deviation at this station is less than 2 $\mu$Gal. The results obtained are summarized in Table 9.

| Date | Gravi | Station | $a_c$ |
|---|---|---|---|
| 12.03.2013 | 209 | Zimm | 3.1 |
| 13.03.2013 | 209 | Zimm | 3.5 |
| 14.03.2013 | 209 | Zimm | 3.4 |
| | | $a_{c\_mean}$ | **3.3** |

Table 9: Proportionality factor evaluated on station Zimm with the instrument 209. The values has been estimated at 3.3 with a standard deviation of the mean of 0.13

*FG5X 209 in the underground laboratory in Walferdange*

The absolute gravity stations of the underground laboratory in Walferdange is used since more than 10 years for conducing regional and international comparisons [25]. During the last key comparison [30] the value of $g$ has been measured on stations *A3*, *B3* and *C3*. The results obtained at different stations in that laboratory are given in Table 10 and Table 11.

| Date | Gravi | Station | $a_c$ |
|---|---|---|---|
| 05.11.2013 | 209 | A3 | 3.7 |
| 06.11.2013 | 209 | B3 | 3.1 |
| 06.11.2013 | 209 | B3 | 3.2 |
| 06.11.2013 | 209 | B3 | 3.3 |
| 07.11.2013 | 209 | C3 | 2.9 |
| 07.11.2013 | 209 | C3 | 3.0 |
| | | $a_{c\_mean}$ | **3.2** |

Table 10: Proportionality factor evaluated on stations A3, B3 and C3 in underground laboratory in Walferdange with the instrument 209. The value has been estimated at 3.2 with a standard deviation of the mean of 0.11.

| Date | Gravi | Station | $a_c$ |
|---|---|---|---|
| 07.11.2013 | 311 | A1 | 2.8 |
| 04.11.2013 | 311 | B5 | 3.5 |
| 05.11.2013 | 311 | B5 | 3.0 |
| 06.11.2013 | 311 | C4 | 3.2 |
| | | $a_{c\_mean}$ | **3.1** |

Table 11: Proportionality factor evaluated on stations A3, B5 and C4 in underground laboratory in Walferdange with the instrument 311. The value has been estimated at 3.1 with a standard deviation of the mean of 0.16.